\documentclass[aps,prc,preprintnumbers,amsmath]{revtex4}

\usepackage{epsfig}
\usepackage{bm}
\usepackage{slashed}
\newcommand{\Tr}{\text{Tr}}
\newcommand{\vect}[1]{\bm{#1}}  
\renewcommand{\vec}{\vect}

\newcommand{\CG}[6]{\left<{#3},{#6}|{#1},{#4};{#2},{#5}\right>}  

\newcommand{\be}{\begin{equation}}
\newcommand{\ee}{\end{equation}}
\newcommand{\bea}{\begin{eqnarray}}
\newcommand{\eea}{\end{eqnarray}}
\newcommand{\bean}{\begin{eqnarray}}
\newcommand{\eean}{\end{eqnarray*}}

\newcommand{\gapproxeq}{\lower
.7ex\hbox{$\;\stackrel{\textstyle >}{\sim}\;$}}
\newcommand{\lapproxeq}{\lower
.7ex\hbox{$\;\stackrel{\textstyle <}{\sim}\;$}}

\def\3bar{$\bar {\hbox{\bf 3}}$}

\newcommand{\cc}{$c\bar{c}$}

\newcommand{\tpn}{^3\mathrm{P}_0}
\newcommand{\tso}{^3\mathrm{S}_1}

\newcommand{\uS}{\textrm{S}}
\newcommand{\uP}{\textrm{P}}

\newcommand{\tpo}{^3\textrm{P}_1}
\newcommand{\tpt}{^3\textrm{P}_2}
\newcommand{\tpj}{^3\textrm{P}_J}

\newcommand{\tdo}{^3\textrm{D}_1}
\newcommand{\tp}{^3\textrm{P}}

\newcommand{\epem}{e^+e^-}

\begin{document}

\title{Charmonium production in $e^+e^- \to \psi +
X(c\bar{c})$ and $\epem \to D^*D_J$}

\author{Frank Close}
\email[E-mail: ]{f.close1@physics.ox.ac.uk}
\affiliation{Rudolf Peierls Centre for Theoretical 
Physics, University of Oxford
,\\ 1 Keble Road, Oxford, OX1 3NP}
 
\author{Clark Downum}
\email[E-mail: ]{c.downum1@physics.ox.ac.uk}
\affiliation{Clarendon Laboratory, University of Oxford,\\ Parks Road, Oxford, 
OX1 3PU}
\date{Version 1.1 - 14 October 2008}
\preprint{OUTP-08-12P}

\begin{abstract}
The dominance of $\chi_0$ in the data for $e^+e^- \to \psi + \chi_J(c\bar{c})$ 
is shown to violate OZI factorization. 
Single gluon exchange gives a non-factorizing effective $\bf{S}\cdot 
\bf{L}$ interaction that generates a large scalar production amplitude. 
This also has observable effects near threshold in $\epem \to {D}^{(*)}D_J$, 
where enhancements of ${D}^*D_0$ and ${D}D_1$ channels are predicted. 
Further tests and implications are discussed.

\end{abstract}

\maketitle

\section{Introduction}
\label{sec:Introduction}

Hadron decays occur dominantly when the colour sources (such as $Q\bar{Q}$) are 
separated by $O(\Lambda_{QCD})$ such that the energy in the fields is enough to 
enable the creation of a light $q\bar{q}$ pair. 
This has been embodied in the phenomenological OZI rule~\cite{OZI}, incorporated
 in models~\cite{models} and more recently studied in lattice QCD~\cite{lattice}
.
It has been shown~\cite{bct} that lattice QCD appears to confirm assumptions 
implicit in many models of strong decays, namely that there is a factorization 
of the constituent spin $S$ and total $J$ of the hadrons (in models this 
equates to a factorization of $S$ and $L$).
Similar conclusions and results are being found in the AdS/QCD 
correspondence \cite{sjb}.
One consequence of factorization is that $\epem \to V+[0^{++}]$ must be smaller 
than at least one of $\epem \to V+[1^{++}]$ and $\epem \to V+[2^{++}]$~
\cite{bct}. This is in marked contrast to data on $\epem \to \psi +X$ where the
$\chi_0$ appears to dominate as reported in 
ref~\cite{ccdata}.  

Decays that involve the creation of heavy flavours where $2m_Q > \Lambda_{QCD}$ 
(e.g. $c\bar{c} \to c\bar{c} + c\bar{c}$) may differ radically from those 
involving creation of light flavours. 
The OZI process is suppressed because $2m_Q > \Lambda_{QCD}$ would require the 
colour fields of force to extend over excessive distances without having 
created light $q\bar{q}$. 
This is highly improbable. 
A way for such decays to be triggered is if the required energy is supplied by a
 hard process such as single gluon exchange. 
As the gluon will in general transmit information about $S$ and $L$ from the 
initial quarks to the created pair, we expect that factorization will not 
occur in such processes. 

Braaten and Lee\cite{braatenlee} present a calculation of 
$\epem \to \psi + \chi_J$
which breaks the limits imposed by 
factorization, and which appears to be supported by the data -- in 
particular the $\chi_0$ dominates the
 $\chi_1$ and $\chi_2$.  This situation has drawn little comment either 
theoretically or experimentally 
despite its important ramifications for strong interaction 
phenomenology.  The purpose of our paper is
to: draw attention to the importance of Braaten and Lee's 
results, expose the origin of their remarkable 
result given their assumptions, extend their result to other cases 
and develop stringent experimental 
tests of the OgE mechanism as opposed to factorized interactions.

In section 2 we review the predictions of factorization. In section 3 we study 
the general structure of the gluon-driven processes producing $\psi + \chi_J$ 
as a function of $J$ and expose 
the origin of the large $\chi_0$ amplitude in ref.~\cite{braatenlee}. 
We apply these results to the production of pairs of charmed mesons and find 
that they lead to an enhancement of $\bar{D}D_1$ and $\bar{D}^*D_0$ channels 
near threshold, but make no contribution to $\bar{D}^*D_1$ or $\bar{D}^*D_2$.
Further, we predict that the helicity amplitude 
$\bar{D}^*(\pm 1)D_2(\mp 2) = 0$,
whereas $\psi(\pm 1)\chi_2(\mp 2)$ is the largest of all the $\psi \chi_2$
amplitudes.
\section{Factorization}
\label{sec:Factorization}

A feature common to many models of hadron decays is that the constituent spins 
factorize from the total angular momentum of the hadrons. 
Ref~\cite{bc} showed that this feature appears to be confirmed by lattice 
QCD~\cite{lattice}, and refs~\cite{bct} showed how this property explains 
relations among various amplitudes that are common to many many specific 
models. 
This factorization property (eq. 5 of ref ~\cite{bct}) underpins
the empirical result that hadron loops 
(such as in \cc~ $\to D\bar{D}, D\bar{D^*}, D^*\bar{D},D^*\bar{D^*} \to$ \cc) 
give universal mass shifts  
throughout an $L$ multiplet such as: $h_c, \chi_0, \chi_1, \chi_2$~\cite{bs}. 
It is also claimed to be consistent with
AdS/QCD correspondence\cite{sjb}

Ref~\cite{bct} derived various consequences of factorization for strong decays 
of hadrons. 
In particular this work showed that factorization constrains the relative 
populations of final states
in $e^+e^- \to V[^3S_1] + (S,A,T)[^3P_J]$.

In the factorization scheme, the decay of a transversely polarized 
$^3S_1 \to ^3S_1 + ^3P_2$, 
with the tensor meson maximally polarized along the decay axis is 
predicted to vanish (see eq. 34 in~\cite{bct}). 
This selection rule is a particular test of factorization. 
Also ref~\cite{bct} found that for $^3S_1$ decay, 
the relative rates in $S$-wave, $S^2$, 
and $D$-wave, $D^2$ are:
 
\be
^3\uS_1+\tpn~:~^3\uS_1+\tpo~:~^3\uS_1+\tpt~:~^1\uS_0 + ^1\uP_1 ~=~ 3S^2~:~4S^2+D
^2~:~6D^2~:~S^2+D^2.
\ee
\noindent Among various consequences, of interest to the present discussion is 
that 
\be
\sigma(^3\uS_1 \to ^3\uS_1+\tpo) =\frac{4}{3}\sigma(^3\uS_1 \to ^3\uS_1+\tpn) 
+ \frac{1}{6}\sigma(^3\uS_1 \to ^3\uS_1+\tpt)
\label{sumrule1}
\ee
\noindent and hence
\be
\sigma(\epem \to ^3\uS_1 \to ^3\uS_1+\tpo) > 
\sigma(\epem \to ^3\uS_1 \to ^3\uS_1+\tpn).
\label{smallscalar}
\ee
\noindent 
One of the central applications of the present paper will be to test these 
predictions against data on $\epem \to \psi + \chi_J$ where preliminary 
indications are that the relation eq.~\ref{smallscalar} is violated
\cite{ccdata}. 


The constraints of factorization become more powerful near threshold where 
$S$-wave dominates.
For a $\tso$ initial state 
\bea
\label{ratioss}
\sigma(^3\uS_1 \to \psi \chi_2) &\to& 0\\ \nonumber
\sigma(^3\uS_1 \to \psi \chi_0) &=& 
\frac{3}{4}\sigma(^3\uS_1 \to \psi \chi_1).
\eea
\noindent 
Analogously, for a $\tdo$ initial state 
\bea
\label{ratiodd}
\sigma(\tdo \to \psi \chi_0)& \to &0\\ \nonumber
\sigma(\tdo \to\psi \chi_1) &=& \frac{5}{3}\sigma(\tdo \to \psi \chi_2).
\eea
\noindent Finally one may allow for a coherent mixture of $\tso$ 
and $\tdo$ initial state. 
Results become model dependent but $\sigma(\psi \chi_0)$ cannot be made larger 
than both  $\sigma(\psi \chi_1)$ and $\sigma(\psi \chi_2)$. 
Thus in the region of threshold factorization forbids a dominant 
$\sigma(\psi \chi_0)$.

This is interesting in view of the data on $e^+e^- \to \psi + X$ at 10.6 GeV 
c.m. energy, which show three prominent enhancements $X$ in $e^+e^- \to 
\psi + X$~\cite{ccdata}, consistent with being the $\eta_c,\eta_c'$ and 
$\chi_0$. 
The observed pattern of states appears radically different to what is seen for
light flavours: the apparent prominence of $\chi_0$ with only a
hint of $\chi_1$ and much suppressed $\chi_2$ contrasts with light flavours 
where $e^+e^- \to \omega f_2$ is clearly seen~\cite{pdg}. 
The charmonium data~\cite{ccdata} are significantly above threshold and so our 
general restrictions against $e^+e^- \to \psi + \chi_0$ need not apply.
However, we shall see that the large rate for this channel is a signal for 
factorization breakdown even away from threshold, and inspires the question: 
what is required to create a dominant $e^+e^- \to \psi + \chi_0$ amplitude?

\section{Gluon exchange structure}
\label{sec:Gluon}

\subsection{Non-relativistic reduction of the Feynman Amplitude}
\label{sec:SpinMatrixElements}

If we require in fig. 1 that the upper \cc~pair produce the $C=-$ meson then 
at leading order four diagrams contribute to $\epem \to \psi \chi_J$. 
These consist of gluon emission from $c$ or $\bar{c}$ in either of two 
topologies: exchange within the $\psi$ 
(figs. 1(a),1(c)) or within the $\chi$ (figs. 1(b),1(d)).  
To contrast with the factorization amplitudes most directly, we shall restrict 
our attention to the threshold region. 
As in ref.~\cite{braatenlee} we set the masses of the $\psi$ and $\chi$ each 
$= 2m_c$. 

We first make a non-relativistic reduction of the Feynman amplitudes into 
2x2 block matrices sandwiched between two-component spinors. 
In so doing it is important to note the role of the (anti-)fermion propagator
between the virtual photon and the gluon. 

Matrix elements in the explicit non-relativistic limit are discussed in a 
consistent phase convention by Ackleh~{\it et al.}~\cite{abs}, 
which we adopt here (see Appendix B of 
ref~\cite{abs}, especially eqs. B5-B7).
Care is required to track phases and
so we define here our choice of some convention dependent quantities.  
Dirac spinors are normalized to 1; 
$\int e^{-i(\vect{p}-\vect{q}).x}d^3x = (2\pi)^3\delta^3(\vect{p} - 
\vect{q})$; $\hbar = c =1$; 
four vectors are denoted by, say, $p$ while three vectors are given by 
$\vect p$; our metric is (+,-,-,-).  
Particle Pauli spinors, $\xi$, and anti-particle 
Pauli spinors, $\eta$ are explicitly:
\begin{align}
    \left(\begin{matrix}
    1\\
    0
  \end{matrix}\right) = \xi_+ = \eta_-\quad
  \left(\begin{matrix}
    0\\
    1
  \end{matrix}\right) = \xi_- = -\eta_+\
\end{align}
The momentum routing is defined in fig. 1.

The analysis of Ackleh~{\it et al.}~was for a $q\bar q$
wavefunction of well defined $^{2S+1}L_J$ 
with no constituent propagator effects considered. 
In $\epem$ annihilation of the present paper, we explicitly consider
the constituent propagator
between the photon and exchanged gluon.  
This has the effect of introducing more than just a single $^{2S+1}L_J$  
state for the photo-produced \cc, i.e. it is not simply $\tso$.
%

From the usual Feynman rules, the matrix element for fig. 1(a) is:
\begin{align}
  \mathcal{M}^\text{a} = \bar{u_1} ig_S\gamma^\nu \frac{i\left(\tfrac{1}{2} 
  \slashed P_1 + \slashed k+m_c\right)}{
  \left(\tfrac{1}{2}P_1+k\right)^2-m_c^2+i\epsilon}
  iee_c\gamma^av_4\frac{-ig_{\nu\mu}}{k^2+i\epsilon}\bar{u_3}ig_S\gamma^\mu v_2.
\end{align}
where $g_S$ is the strong coupling constant and $e_c$ is the 
ratio of the electron
and $c$ quark charges and $e$ is the electron charge.
Our primary interest is in the $J$-dependence of the production 
of $\tso + \tpj$ near 
threshold. 

In NRQCD the appropriate $L$ state is extracted by expanding 
the matrix element into a double Taylor series of the internal momenta
$q_1$ and $q_2$ (about 0) and selecting the appropriate power of the 
internal momenta.  The angular dependence of the appropriate term in 
the Taylor series is recombined with the spin of the hadron using covariant 
generalizations of Clebsh-Gordon coeffecients and the radial dependence 
is absorbed into a model-dependent vacuum saturated analogues of the 
NRQCD matrix element, 
$\langle {O}_1\rangle_\psi$ say.  The $\langle {O}_1\rangle_{\tpj}$ are
approximately independent of $J$, as one would expect from any spherically
symmetric hadronization process, and as such are collectively denoted
$\langle {O}_1\rangle_{\tp}$.

For the S and P-wave mesons of interest, the appropriate term is the one 
with a single power of $q_2$ and zero powers of $q_1$ which is ensured 
by setting $q_1=0$.  Since we are interested in the threshold limit, 
we simplify the series by taking $\vect P_1 = \vect P_2 = \vect 0$.  After 
expansion and algebraic simplification of the matrix element, the $q_2$ 
will be replaced with $\vect{\varepsilon^*}_L$ and its radial dependence
absorbed into the vacuum saturated matrix element.   Finally the spin 
of the $\tpj$ and its orbital angular momentum will be combined to obtain 
the appropriate $J$.  This is the non-relativistic reduction of the 
covariant NRQCD technique for $\tso + \tpj$.  

The matrix element is effectively simplified to:
\begin{equation}
   \mathcal{M}^\text{a} =
  \underbrace{-ig_S^2ee_c\frac{8}{s^2}
\left(\sqrt{\frac{E+m_c}{2m_c}}\right)^2}_N
  \xi_1\dagger 
  \begin{pmatrix}
    1\\
    0
  \end{pmatrix}^\dagger\gamma^0\gamma^\nu
  \left(\tfrac{1}{2}\slashed P_1 + \slashed k +m \right)\gamma^a
  \begin{pmatrix}
    -\vect{\sigma}.\hat q_2\\
    1
  \end{pmatrix}\eta_2\xi_2^\dagger 
  \begin{pmatrix}
    1\\
    \vect{\sigma}.\hat q_2
  \end{pmatrix}^\dagger\gamma_\nu 
  \begin{pmatrix}
    0\\
    1
  \end{pmatrix}\eta_1\nonumber
\end{equation}
where $\hat q_2 \equiv \vect q_2/2m_c$
  
We generalise the above to allow the quarks of the flavour produced by the 
gluon to have mass $m_f$. Then we
define  $\delta = 2m_f/m_c$; in the particular case of
$\psi+\chi_J$, $m_f = m_c$ and hence $\delta =2$. For flavoured mesons
with $m_f \to 0$ one has $\delta \to 0$. We shall
retain the general form so that applications to other 
combinations of flavours may be made.
The $\hat q_2 \equiv \vect q_2/2m_c$
then generalises trivially to contain the relevant
$m_{f}$ in the denominator: 
$\hat q_{2f} \equiv \vect q_2/2m_f = \tfrac{2}{\delta} \hat q_2$.
$N$ also has an implicit dependence on $\delta$ but the dependence does 
not matter for the relative rates we are interested in.   
Hence
  \begin{align}
    \mathcal{M}^\text{a}
&=N\xi_1^\dagger 
  \begin{pmatrix}
    1\\
    0
  \end{pmatrix}^\dagger
  \gamma^0 \gamma^\nu 
  \left( m_c\gamma^0 - \tfrac{1}{2}\vect{P_1}. \right.
  \vect{\gamma} + m_c\delta\gamma^0 - \vect{k}.\vect{\gamma} + m_c 
  \left. \right)
  \begin{pmatrix}
    0&\sigma^a\\
    -\sigma^a & 0
  \end{pmatrix}
  \begin{pmatrix}
    -\vect{\sigma}.\hat q_2 \\
    1
  \end{pmatrix}
  \eta_4\xi_3^\dagger\nonumber\\
  &\hspace{.5in}
  \begin{pmatrix}
    1\\
    \vect{\sigma}.\hat q_{2f} 
  \end{pmatrix}^\dagger\gamma^0\gamma_\nu
  \begin{pmatrix}
    0\\
    1
  \end{pmatrix}\eta_2
\end{align}
We can split the equation into a time-like component ($\gamma^0\gamma^0$) and a 
space-like component ($\vec{\gamma}\vec{\gamma}$).  
Taking the leading order terms for the time-like component,
\begin{align}
 \mathcal{M}^{\text{a}\,\gamma^0\gamma^0}
 &= N \Tr\left\{\vphantom{\frac{1}{2}}\right.
    \begin{pmatrix}
    1\\
    0
  \end{pmatrix}^\dagger
  \begin{pmatrix}
    m_c(2+\delta)&-\vect{\sigma}.\vect{q}_2\\
    \vect{\sigma}.\vect{q}_2 & -m_c\delta
  \end{pmatrix}
  \begin{pmatrix}
    0&\sigma^a\\
    -\sigma^a & 0
  \end{pmatrix}
  \begin{pmatrix}
    -\vect{\sigma}.\hat{q}_2\\
    1
  \end{pmatrix}
  \left(\eta_4\xi_3^\dagger\right) \nonumber\\
&\quad\quad
  \begin{pmatrix}
    1\\
    \vect{\sigma}.\hat{q}_{2f}
  \end{pmatrix}^\dagger
  \begin{pmatrix}
    0\\
    1
  \end{pmatrix}\left(\eta_2\xi_1^\dagger\right)
  \left. \vphantom{\frac{1}{2}} \right\} 
\end{align}

Having written the matrix element in terms of block 2x2 matrices and particle, 
anti-particle Pauli spinors ($\xi,\eta$),
we can use 
\begin{align}
  \sum_{\lambda_1,\lambda_2}
\CG{\tfrac{1}{2}}{\tfrac{1}{2}}{1}{\bar \lambda}{\lambda}{S}\eta_{\bar \lambda}
 \xi_{\lambda}^\dagger = -\frac{1}{\sqrt{2}} 
\vec\sigma.\vec{\varepsilon}^*\left( S \right)
\end{align}
whereby the substitution, 
$\eta\xi^\dagger \mapsto -\tfrac{1}{\sqrt{2}} \vec \sigma . 
\vec{\varepsilon}^*$ projects the matrix element into the spin triplet state.
The terms of exactly one power of $\vect{q_2}$ are retained 
($\vect q_1=0$ ensures that 
only terms with zero powers of $\vect q_1$ were kept).  $\vect q_2$ 
is replaced by
$\vect{\varepsilon^*}_L$ and its radial dependence absorbed 
into the vacuum saturated 
analogues of the NRQCD matrix elements.
The result is the projection of the matrix element into the $\tso + \tp$ state:
\begin{align}
\mathcal{M}^{\text{a}\,\gamma^0\gamma^0}(\gamma^*\to \tso + \tp)
&= \frac{1}{2} N\frac{2+\delta}{2}\frac{2}{\delta}
\langle {O}_1\rangle_{\tso}\langle {O}_1\rangle_{\tp}\Tr
\left\{\vphantom{\frac{1}{2}}  \right.
\vect{\sigma}.\vect{a} 
\vect{\sigma}.\vect{\varepsilon_2}^* 
\vect{\sigma}.\vect{\varepsilon_L}^* 
\vect{\sigma}.\vect{\varepsilon_1}^*
\left.\vphantom{\frac{1}{2}}\right\}\nonumber
\end{align}
and then using  
\begin{equation}
\tfrac{1}{2}\Tr\left\{\sigma^i\sigma^j\sigma^k\sigma^l \right\} = 
    \delta^{ij}\delta^{kl} - \delta^{ik} \delta^{jl} + \delta^{il} \delta^{jk}
\label{pauli1}
\end{equation} 
we obtain
\begin{align}
&\mathcal{M}^{\text{a}\,\gamma^0\gamma^0}(\gamma^*\to \tso + \tp)=
N\frac{2+\delta}{2} \frac{2}{\delta}
\langle {O}_1\rangle_{\tso}\langle {O}_1\rangle_{\tp}
\left( \right.
\vect{\varepsilon_2}^*.\vect{a} 
\vect{\varepsilon_L}^*.\vect{\varepsilon_1}^* 
- \vect{\varepsilon_L}^*.\vect{a} 
\vect{\varepsilon_2}^*.\vect{\varepsilon_1}^* 
+ \vect{\varepsilon_1}^*.\vect{a} 
\vect{\varepsilon_2}^*.\vect{\varepsilon_L}^*
\left.\right).
\end{align}
where
$\vect{a}$ is the polarisation of the photon, $\vect{n}$ is the 
polarisation of the spatial components of the gluon, 
$\vect \varepsilon_L^*$ is the orbital angular momentum of the $\tp$ meson
and $\vect \varepsilon_1^*,\vect\varepsilon_2^*$ are the 
spin polarisation tensors 
of the $\tso , \tp$ mesons respectively.  The time-like components of the 
amplitudes cancel among the various graphs.  Therefore we do not consider
them further and focus on the space-like components.

For the space-like component, we have
\begin{align}
   \mathcal{M}^{\text{a}\,\vect\gamma\vect\gamma}
&= - N \Tr\left\{
    \begin{pmatrix}
    1\\
    0
  \end{pmatrix}^\dagger
  \begin{pmatrix}
    0&\sigma^n\\
    +\sigma^n & 0
  \end{pmatrix}  
  \begin{pmatrix}
    m_c(2+\delta)&-\vect{\sigma}.\vect{q}\\
    \vect{\sigma}.\vect{q} & -{m_c}\delta
  \end{pmatrix}
  \begin{pmatrix}
    0&\sigma^a\\
    -\sigma^a & 0
  \end{pmatrix}
  \nonumber\right.\\
  &\hspace{.5in}\left.\times
  \begin{pmatrix}
    -\vect{\sigma}.\hat{q_2}\\
    1
  \end{pmatrix}
  \left(\eta_4\xi_3^\dagger\right)
  \begin{pmatrix}
    1\\
    \vect{\sigma}.\hat q_{2f} 
  \end{pmatrix}^\dagger
  \begin{pmatrix}
    0&\sigma^n\\
    +\sigma^n & 0
  \end{pmatrix}  
  \begin{pmatrix}
    0\\	
    1
  \end{pmatrix}\left(\eta_2\xi_1^\dagger\right)
\right\} \nonumber\\
\Rightarrow\mathcal{M}^{\text{a}\,\vect\gamma\vect\gamma}
(\gamma^*\to \tso + \tp)
& = -\frac{1}{2}N \langle {O}_1\rangle_{\tso}\langle {O}_1\rangle_{\tp} \Tr
\left\{\vphantom{\frac{1}{2}} \right.
\left(\vect{\sigma}. \right.
\vect{\varepsilon_L}^*\vect{\sigma}.\vect{a}\ -  \frac{\delta}{2}
\vect{\sigma}.\vect{a}\vect{\sigma}.\vect{\varepsilon_L}^*
\left.\right)\vect{\sigma}
.\vect{\varepsilon_2}^* \vect{\sigma}.\vect{\varepsilon_1}^*
\left.\vphantom{\frac{1}{2}}\right\}.\label{ogepsi} 
\end{align}
Then as before, using eq.\ref{pauli1}, this may be written
\begin{align}
&\mathcal{M}^{\text{a}\,\vec\gamma\vec\gamma}(\gamma^*\to \tso + \tp)=N
\langle {O}_1\rangle_{\tso}\langle {O}_1\rangle_{\tp}
\left(\vphantom{\frac{1}{2}}\right.
\frac{\delta+2}{2}\vect{\varepsilon_1}^*.\vect{a} \vect{\varepsilon_L}^*.
\vect{\varepsilon_2}^* - \frac{\delta - 2}{2}\vect{\varepsilon_L}^*.
\vect{a} \vect{\varepsilon_2}^*.\vect{\varepsilon_1}^* - 
\frac{\delta+2}{2}\vect{\varepsilon_2}^*.\vect{a} 
\vect{\varepsilon_L}^*.\vect{\varepsilon_1}^*
\left.\vphantom{\frac{1}{2}}\right).
\end{align}

The amplitude in fig. 1(b) is calculated in similar fashion and gives
\begin{align} 
\mathcal{M}^{\text{b}\,\vec\gamma\vec\gamma}(\gamma^*\to \tso + \tp)
&= \frac{1}{2}N
\frac{2+\delta}{2} \langle {O}_1\rangle_{\tso}\langle {O}_1\rangle_{\tp} 
\Tr
\left\{ \vphantom{\frac{1}{2}} \right.
\vect{\sigma}.\vect{a} 
\vect{\sigma}^{n} \vect{\sigma}.\vect{\varepsilon_L}^*
\vect{\sigma}.\vect{\varepsilon_2}^* \vect{\sigma}_{n} 
\vect{\sigma}.\vect{\varepsilon_1}^*
\left. \vphantom{\frac{1}{2}} \right\}\label{ogechi}\\
&=N\frac{2+\delta}{2} \langle {O}_1\rangle_{\tso}\langle {O}_1\rangle_{\tp}
\left(\vphantom{\frac{1}{2}}\right.
3 \vect{\varepsilon_1}^*.\vect{a} 
\vect{\varepsilon_L}^*.\vect{\varepsilon_2}^*
+  \vect{\varepsilon_2}^*.\vect{a} 
\vect{\varepsilon_L}^*.\vect{\varepsilon_1}^*
-  \vect{\varepsilon_L}^*.\vect{a}
\vect{\varepsilon_2}^*.\vect{\varepsilon_1}^*
\left.\vphantom{\frac{1}{2}}\right)\nonumber
\end{align}

Explicit calculation of the other diagrams confirms the symmetries:

\begin{align}
 \mathcal{M}^{\text{a}\,\gamma^0\gamma^0}(\gamma^*\to \tso + \tp) 
& =  \mathcal{M}^{\text{c}\,\gamma^0\gamma^0}(\gamma^*\to \tso + \tp)\nonumber\\
 &=  - \mathcal{M}^{\text{b}\,\gamma^0\gamma^0}(\gamma^*\to \tso + \tp)
 = - \mathcal{M}^{\text{d}\,\gamma^0\gamma^0}(\gamma^*\to \tso + \tp) 
\nonumber\\
 \mathcal{M}^{\text{a}\,\vec\gamma\vec\gamma}(\gamma^*\to \tso + \tp) 
 & =  \mathcal{M}^{\text{c}\,\vec\gamma\vec\gamma}
 (\gamma^*\to \tso + \tp);\nonumber\\
 \mathcal{M}^{\text{b}\,\vec\gamma\vec\gamma}(\gamma^*\to \tso + \tp)
 &= \mathcal{M}^{\text{d}\,\vec\gamma\vec\gamma}(\gamma^*\to \tso + \tp)
\end{align}

Thus the total amplitude at threshold becomes
\begin{align}
  \mathcal{M}(\gamma^*\to \tso + \tp)&=N
  \langle {O}_1\rangle_{\tso}\langle {O}_1\rangle_{\tp}
  \left(\vphantom{\frac{1}{2}} \right.
  (2\delta+4)\vect{\varepsilon_1}^*.\vect{a} 
  \vect{\varepsilon_L}^*.\vect{\varepsilon_2}^* 
  - \delta\vect{\varepsilon_L}^*.\vect{a} 
  \vect{\varepsilon_1}^*.\vect{\varepsilon_2}^* 
  \left. \vphantom{\frac{1}{2}} \right)
  \label{totalamp}
\end{align}

For arbitrary final state meson momentum the amplitude is
\begin{align}
  \mathcal{M}(\gamma^{*\alpha}\to \tso + \tp )
&\propto
4m_c^2\,\varepsilon_1\cdot \varepsilon_2 (6P_1^\alpha P_1\cdot \varepsilon_L 
                                    - s\varepsilon_L^\alpha)\nonumber\\
&+\varepsilon_2\cdot \varepsilon_L s 
(s\varepsilon_1^\alpha-2P_1^\alpha \varepsilon_1\cdot P_2) 
+24 m_c^2\, P_1\cdot \varepsilon_L ( 
  \varepsilon_2^\alpha \varepsilon_1\cdot P_2 
 -\varepsilon_1^\alpha \varepsilon_2\cdot P_1
 ).
\label{jungilamp}
\end{align}
$\alpha$ is the photon's helicity.
We are grateful to Dr Jungil Lee for providing this expression.
As in the threshold limit, this indeed reduces to our 
eq.~\ref{totalamp} with $\delta =2$ 
for physical values of the photon helicity $\alpha$.

Of the two contributions in eq.~\ref{totalamp}, 
only the $\vect{\varepsilon_L}^*.\vect{a} 
\vect{\varepsilon_1}^*.\vect{\varepsilon_2}^* $ factorizes $L$ and $S$.
The first term couples $L_{\tpj}$ and $S_{\tpj}$ and survives everywhere 
except the
unphysical case of threshold for $\delta = -2$. 
In the physical cases of interest, this term
turns out to dominate.

\subsection{Amplitudes in NRQCD}
\label{sec:NRQCD}

In the particular limit of $\psi \chi$ at threshold, 
$\delta=2$ the amplitude is:
\begin{align}
  \mathcal{M}(\gamma^*\to \psi + c\bar c (\tp))&=2N
  \langle {O}_1\rangle_{\tso}\langle {O}_1\rangle_{\tp}
  \left( \right.
  4\vect{\varepsilon_1}^*.\vect{a} 
  \vect{\varepsilon_L}^*.\vect{\varepsilon_2}^* 
  - \vect{\varepsilon_L}^*.\vect{a} 
  \vect{\varepsilon_1}^*.\vect{\varepsilon_2}^* 
  \left. \right).
  \label{eq:feynstructure}
\end{align}
For charm pair production at threshold where $\delta=0$ the
amplitude simplifies to:
\begin{align}
  \mathcal{M}(\gamma^* \to D^* + c\overline{q} (\tp))&=N
  \langle {O}_1\rangle_{\tso}\langle {O}_1\rangle_{\tp}
  \left( \right.
  4\vect{\varepsilon_1}^*.\vect{a} 
  \vect{\varepsilon_L}^*.\vect{\varepsilon_2}^* 
  \left. \right).
  \label{eq:feynstructure2}
\end{align}
Note that $N$ and the vacuum saturated matrix elements
will be different in the charm pair case as opposed to the double
charmonium case, 
but this will not matter for relative amplitudes in the 
$D^*D_J$ channels of interest.

Eq.~\ref{jungilamp} is the starting point from which the 
amplitudes of ref~\cite{braatenlee}
can be obtained. Our Pauli decomposition at threshold, eq.~\ref{totalamp}
reveals the origin of some particularly interesting results. 
In particular, for later reference, we draw attention to the
NR reduction of the amplitude of the topology with OgE within the 
$\psi$ (figs. 1(a) and 1(b),
equation \ref{ogepsi}), which involves the momentum $\vect{q}_2$ 
flowing through the 
photon-\cc~vertex and along the virtual fermion line connecting photon to 
gluon. 
The photo-produced \cc~pair is then clearly not in a simple $\tso$ state.
Physically, contributions other than $\tso$ would vanish if the amplitude were 
proportional only to wavefunction at the origin; it is the 
spatial propagation 
away from the `origin', associated with the propagator, that enables the 
non-zero amplitudes associated with these other configurations at threshold.

First consider the factorizing contribution in eq.~\ref{totalamp}, namely
$\vect{\varepsilon_L}^*.\vect{a} 
\vect{\varepsilon_1}^*.\vect{\varepsilon_2}^* $. The relative size of 
amplitudes arising from
this term alone 
after combining with suitable Clebsch-Gordan coefficients 
in order to give helicity amplitudes for various $\chi_J$ production in 
association with $\psi$ are as follows.

\begin{align}
\psi(-)\chi_2(++) = 1;  \psi(0)\chi_2(+) = 
1/\sqrt{2}; & \psi(+)\chi_2(0) = 1/\sqrt{6}\\
 \psi(0)\chi_1(+) = 1/\sqrt{2}; & \psi(+)\chi_1(0) = 1/\sqrt{2}\\
 &\psi(+)\chi_0(0) = 1/\sqrt{3}
\end{align}
Thus for the unphysical case $\delta = -1/2$ where 
this term alone is present, one finds
a result consistent with factorization as expected:

\begin{equation}
\sigma(\psi\chi_2):\sigma(\psi\chi_1):\sigma(\psi\chi_0) = 5:3:1
\end{equation}

In the particular limit of threshold and $\delta =0$, only the 
``non-factorizing" contribution  
$ \vect{\varepsilon_2}^*.\vect{\varepsilon_L}^*
\equiv {\bf S_{\chi}}\cdot{\bf L_{\chi}}$
survives. This limit can be realised if the produced quarks 
have masses $\to 0$, as in the production
of charmed mesons, $D^*, D_J$.

The expectation values of this term, after combining with the 
Clebsch-Gordan coefficients for $L \times S \to J$ 
appropriate for $\chi_J$ give, in the same normalisation as above 
\[
 \psi(+)\chi_J(0) = [\langle J0 | 11;1-1 \rangle + \langle J0 | 1-1;11 
 \rangle - \langle J0 | 10;10 \rangle ]
 \]
which vanishes for $J=1,2$ and is non-zero for $J=0$ where the constructive
interference of the three terms gives
the large value of $\sqrt{3}$. 

For the case of $\epem \to \psi + \chi_J$ the amplitude is given 
by eq.~\ref{eq:feynstructure}.
The non-factorizing
$ \vect{
\varepsilon_2}^*.\vect{\varepsilon_L}^*\equiv {\bf S_{\chi}}\cdot{\bf L_{\chi}}$
 term dominates by a factor of 4 relative to $ \vect{\varepsilon_1}^*.
\vect{\varepsilon_2}^*\equiv {\bf S_{\chi}}
\cdot {\bf S_{\psi}}$. While the factorizing term alone 
gave amplitude $1/\sqrt{3}$, 
the combination of
the two terms now gives $\tfrac{1}{\sqrt{3}}(1 -12)$.
Thus in leading order of NRQCD at threshold, 
we find the following relative rates: 
 \be
 \sigma(\epem \to \psi[\chi_2:\chi_1;\chi_0]) = 5:3:121.
 \label{bigscalar}
 \ee
This agrees with the result in ref~\cite{braatenlee}; (to extract 
the OgE contribution, set $Y =0$ in the amplitudes eqs. A3 in appendix A of 
ref~\cite{braatenlee}).
  
The same combination of 
Clebsch-Gordan coefficients arises in the longitudinal amplitude for 
$ \epem \to \psi(0)\chi_J(0)$.
Thus the origin of the large scalar amplitude at threshold is due to 
a $\bf{S_\chi}\cdot \bf{L_\chi}$ transition 
operator between the initial state and the $L=1$ $\chi_J$ state which vanishes 
for all but $J=0$.

Note that the $\psi \chi_1$ and $\psi \chi_2$ amplitudes at 
threshold factorize.
If the $\psi \chi_2$ amplitudes can be isolated, there is an interesting test
concerning the non-zero $\psi(-)\chi_2(++)$.  
In ref~\cite{bct} (table III and eqs. $36-41$),  
it was noted for a $\tso$ initial state that factorization implies 
that the $\psi(-)\chi_2(++)$ amplitude vanishes in 
S-wave, and that the D-wave amplitudes destructively cancel. 
In OgE however, there is more than simply $\tso$ in the initial state
with the result that the S-wave is non-zero
and as noted above we find
\be
a[\psi(-)\chi_2(++)]:a[\psi(0)\chi_2(+)]:a[\psi(+)\chi_2(0)] = 
1:1/\sqrt{2}:1/\sqrt{6}
\label{vtratios}
\ee
in accord with S-wave dominance.

Thus the $\psi(-)\chi_2(++)$ amplitude, far from vanishing, is predicted to be 
the dominant helicity state for $\psi\chi_2$. This provides an interesting 
test for the presence of non $^3S_1$ contributions in $\epem$.



\subsection{One gluon exchange amplitude for arbitrary momentum}
\label{subsec:braaten}

Eq.~\ref{jungilamp} is the amplitude for arbitrary $r$.
From these amplitudes ref \cite{braatenlee} obtains rates 
for $\psi(\lambda_1) + \chi_J(\lambda_2)$, where $\lambda_{1,2}
$ are the helicities of the charmonium states, as a function of 
$r^2 \equiv 16m_c^2/s$. 
At the 10.6 GeV c.m. energy of the data~\cite{ccdata}, where 
$r^2 \equiv 16m_c^2/s = 0.28$, 
ref~\cite{braatenlee} finds for the OgE contribution to the cross sections 
\be
\sigma(\psi \chi_2:\psi\chi_1:\psi\chi_0) 
\sim 3:2:12.
\ee

The $L$ and $S$ dependence factorizes in three of the terms; 
the only term coupling $L$ and $S$ is the one
already identified.
Compared to the results at threshold, the relative sizes of 
$\sigma(\psi \chi_2):\sigma(\psi\chi_1)$ have not changed much but 
there is an order of magnitude 
relative reduction of $\sigma(\psi \chi_0)$ at the higher energy. Nonetheless,
it is the non-factorizing term that continues to dominate. 
Relatively large scalar meson production is predicted as 
a robust phenomenon at all energies.


\subsection{Charm pair production in NRQCD}
\label{sec:charmpair}

The above analysis can be applied to charm pair production, $\epem \to D_J \bar{
D}^{(*)}$ or $\bar{D}_JD^{(*)}$.
Near threshold, factorization predicts~\cite{bct} that the ratios of 
cross-sections from initial $^3S_1, ^3D_1$ or vector hybrid to the states 
$D^*D_2:D^*D_{1}(^3P):D^*D_0$ are as shown in table \ref{ozicharm}.

\begin{table} 
  \begin{tabular}{l@{$\quad$}cccc}
   \hline
           &  $D^*D_2\;$&$D^*D_{1}(^3P)\;$&$D^*D_0\;$\\
   \hline
   $^3S_1$ & 0&4&3 \\
   $^3D_1$ & 3&5&0 \\
   ~hybrid & 6&3&4 \\
   \hline
 \end{tabular}
 \caption{Relative sizes of transversely polarized decay amplitudes from OZI in 
charm pair production, $\delta\rightarrow 0$.}
   \label{ozicharm}
\end{table}

As in the \cc~+\cc~ case we see here too that OZI factorization requires that 
the scalar production cannot be larger than both the axial and tensor channels.

The interesting feature for charm pair production is that the OZI, or 
``flux-tube breaking'', mechanism is dynamically allowed, but that OgE 
can also be anticipated to be present.
The relative sizes of these OgE contributions to the various charmed 
meson channels are calculated analogously to before, except that now we have $
\delta \to 0$.  
Within the approximation that the large $Q^2$ photon produces the $c\bar c$, and
 the gluon then produces $q\bar q$, the matrix element for $D^*D_J$ is given by 
eq.~\ref{eq:feynstructure2}.
In practice there will also be some contribution where $\gamma^* \rightarrow q
\bar q$ followed by $g\rightarrow c\bar c$.  
This is generally expected to be small because of unfavorable energetics. 

It is interesting to note that the $\vect \varepsilon^*_1 . \vect a, 
\vect\varepsilon^*_L . \vect\varepsilon^*_2$ structure of 
eq.~\ref{eq:feynstructure2} is the same 
structure that, in eq.~\ref{eq:feynstructure}, was responsible for 
the large $\psi \chi_0$ amplitude in double charmonium production.  
Within the above approximation for $D^*D_J$ production, we see that the 
$D^*D_0$ channel is the only one driven by OgE, a maximal violation of 
factorization.  
The $DD_1(^1P_1)$ channel is also non-zero.

The double charmonium production is in practice the 
``worst case" for the scalar dominance.
Yet it is clearly visible in the data, and our analysis at threshold shows why.
For charmed meson production, where $g \to q\bar{q}$ with light flavours and
 $\delta \to 0$,
the effect will necessarily be bigger, effectively infinite 
for reasonable parametrisation of
$m_q/m_c$.

The effective absence of OgE contributions for $\epem \to D^*D_1$ and $D^*D_2$, 
which becomes exact in the limit $\delta \to 0$,
implies that the factorization selection rules should be 
particularly robust here.
Thus we predict that at threshold, the helicity amplitude 
$\bar{D}^*(\pm 1)D_2(\mp 2) = 0$,
in contrast to the case of $\psi(\pm 1)\chi_2(\mp 2)$ where it is the 
largest of all the $\psi \chi_2$
amplitudes.

The OgE selection of $D^*D_0$ and $DD_1$ in the threshold region may play 
some role in generating the enhancement seen as $Y(4260)$~\cite{4260} (where 
charmed mesons in a relative $S$ wave might rescatter to form $\psi\pi\pi$).  
We shall return to the $\gamma^* \rightarrow q
\bar q$ process in more detail elsewhere.
 
\section{Conclusions}
 
The intrigue of $\epem \to \psi + X$ can hardly be overstated. 
There is a clear spectrum of $C=+$ charmonium states, whose pattern is not yet 
explained and in addition an enigmatic structure around 3940 MeV. 
As this is in the mass region where $C=+$ charmonium hybrids could lurk~
\cite{cchybrids}, any theoretical modelling of this requires first 
understanding the population of the other bumps.
The $\chi_J$ production in particular needs to be understood; it seems 
empirically dominated by $\chi_0$, at least when a mass 
fit is made to the cross section;
 no $J^P$ analysis has been made.
We have drawn attention to the fact that the dominance of $\chi_0$ contrasts 
radically with normal OZI expectation where $\chi_0$ would be relatively small. 
We demonstrated that OgE gives large scalar, confirming Braaten and Lee 
\cite{braatenlee}, and have exposed the origin and significance of this result 
which had not previously been recognised as a sharp test of dynamics.
 
Thus if dominance of $\psi \chi_0$ is confirmed over a range of $q^2$ away from 
threshold, this would support OgE as the dominant decay mechanism.
Conversely, if data near threshold confirm $a[V(-)T(++)] \to 0$, this would 
signal factorization being dominant. 
In any event, we anticipate that the relative populations and helicity 
structures of $\psi \chi_J$ will vary with $q^2$. 
We recommend that this be investigated in $\epem$ annihilation at super-B 
factories by means of ISR to access a range of energies. 
In particular experiment should attempt to measure the spin dependence of $\epem
 \to \psi \chi_2$ as a function of $q^2$ and compare with the analogous 
 amplitudes in $e^+e^- \to \omega f_2$ or $e^+e^- \to \rho f_2$.

When applied to charm pairs we find violation of OZI rules here too: OgE 
selects the $D^*D_0$ and $DD_1$ channels near threshold. 
The charm pair arena is interesting as it potentially enables us to test the 
relative role of OgE versus OZI dynamics.
It also shows a possible mechanism for generating enhancements in these $S$-wave
 production channels which, by constituent rearrangement, may produce these 
 structures, such as seen in $\epem \to \psi \pi\pi$ at 4260 MeV~\cite{4260}.

Our approach has demonstrated the breakdown of factorization in OgE.  
Ackleh,~{\it et. al.} implicitly noted this in ref~\cite{abs} 
at least for on-shell 
constituents with a definite $^{2S+1}L_J$ initial state.  
Our work goes beyond this by exposing the important role propagator effects can
play; in particular we have found this to be essential in matching to the 
NRQCD work of ref ~\cite{braatenlee}, at least at threshold.

As a result, we have exposed the origin of the large amplitude for $\psi\chi_0$ 
relative to $\psi\chi_{1,2}$ which manifestly violates factorization and for
which there are preliminary hints in data.
We urge experiment to use spin analysis to confront this and other $J^P$ tests 
in both charm pair and light hadron production in order to sharpen 
understanding of the relative importance of OZI and non-factorizing dynamics.

\section*{Acknowledgements}

We are indebted to Jungil Lee for discussions on the NRQCD amplitudes and in 
comparing our amplitudes
with those of ref~\cite{braatenlee}.

This work is supported by grants from the Science \& Technology Facilities 
Council (UK) and in part by the EU Contract No. MRTN-CT-2006-035482, 
``FLAVIAnet.''



\appendix


\newpage

\begin{figure}[h]
  \includegraphics{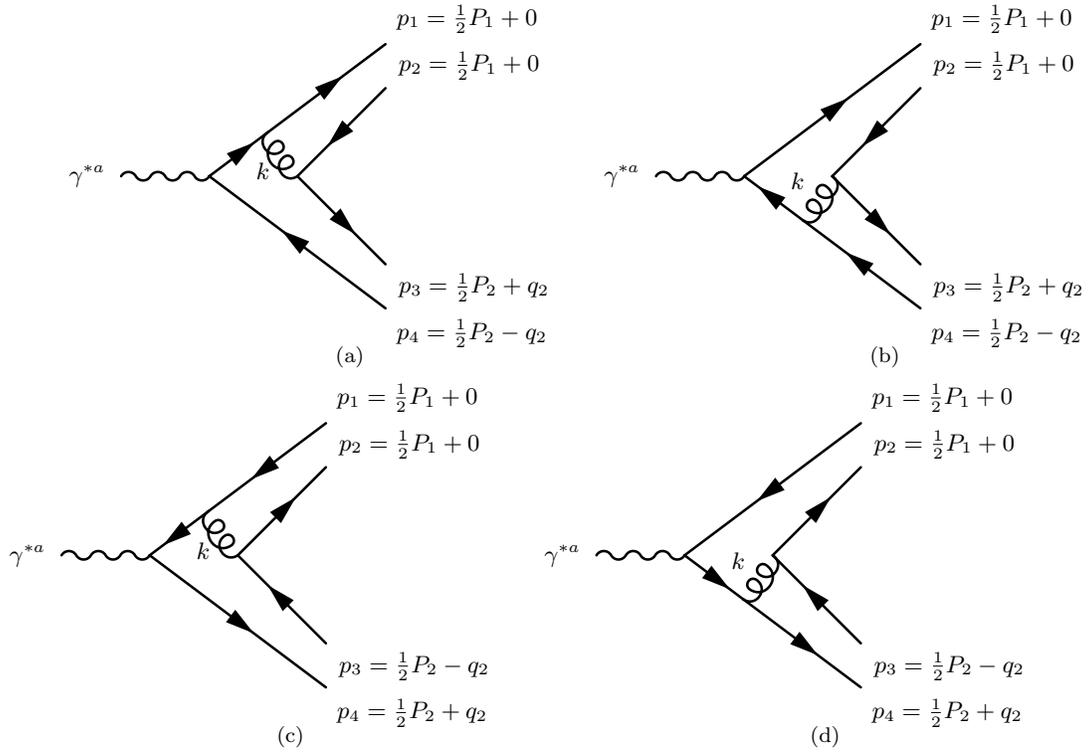}
  \caption{The four topologies of the OgE model for 
$\epem \rightarrow c\bar cc\bar c$. }
\end{figure}

\end{document}